\documentclass[11pt]{article}

\usepackage{amsmath}
\usepackage{graphicx}
\usepackage{amsfonts}
\usepackage{amssymb}
\usepackage{epsfig}
\usepackage{color}
\usepackage{psfrag}
\usepackage{epstopdf}

\usepackage{caption}
\usepackage{subcaption}

\newcommand{\EQ}{\begin{equation}}
\newcommand{\EN}{\end{equation}}
\newcommand{\be}{\begin{equation}}
\newcommand{\ee}{\end{equation}}
\newcommand{\bea}{\begin{eqnarray}}
\newcommand{\eea}{\end{eqnarray}}

\DeclareMathOperator*{\SumInt}{%
\mathchoice%
  {\ooalign{$\displaystyle\sum$\cr\hidewidth$\displaystyle\int$\hidewidth\cr}}
  {\ooalign{\raisebox{.14\height}{\scalebox{.7}{$\textstyle\sum$}}\cr\hidewidth$\textstyle\int$\hidewidth\cr}}
  {\ooalign{\raisebox{.2\height}{\scalebox{.6}{$\scriptstyle\sum$}}\cr$\scriptstyle\int$\cr}}
  {\ooalign{\raisebox{.2\height}{\scalebox{.6}{$\scriptstyle\sum$}}\cr$\scriptstyle\int$\cr}}
}

\begin{document} \setcounter{page}{0}
\topmargin 0pt
\oddsidemargin 5mm
\renewcommand{\thefootnote}{\arabic{footnote}}
\newpage
\setcounter{page}{0}
\topmargin 0pt
\oddsidemargin 5mm
\renewcommand{\thefootnote}{\arabic{footnote}}
\newpage
\begin{titlepage}
\begin{flushright}
%SISSA 40/2012/EP \\
%DFTT 9/2007
\end{flushright}
\vspace{0.5cm}
\begin{center}
{\large {\bf Persistent oscillations after quantum quenches:\\ The inhomogeneous case}}\\
%{\bf  of random cluster and $O(n)$ models}}\\
\vspace{1.8cm}
{\large Gesualdo Delfino$^{1,2}$}\\
\vspace{0.5cm}
{\em $^1$SISSA -- Via Bonomea 265, 34136 Trieste, Italy}\\
{\em $^2$INFN sezione di Trieste, 34100 Trieste, Italy}\\
%{\em E-mail: delfino@sissa.it}\\
%\vspace{0.5cm}
%{\large and}\\
%\vspace{0.5cm}
%{\large P. Simonetti}\\
%\vspace{0.5cm}
%{\em Department of Physics, University of Wales Swansea,\\
%Singleton Park, Swansea SA2 8PP, United Kingdom}\\
%{\em email: p.simonetti@swansea.ac.uk}\\
\end{center}
\vspace{1.2cm}

\renewcommand{\thefootnote}{\arabic{footnote}}
\setcounter{footnote}{0}

\begin{abstract}
\noindent
We previously showed that a quantum quench in a one-dimensional translation invariant system produces undamped oscillations of a local observable when the post-quench state includes a single-quasiparticle mode and the observable couples to that mode [J.	~Phys.~A ~47 (2014)~402001]. Here we consider quenches that break initial translation invariance. Focusing on quenches performed only on an interval of the whole system, we analytically determine the time evolution of local observables, which occurs inside a truncated light cone spreading away from the quenched interval as time increases. If the quench excites a single-quasiparticle mode, oscillations with the frequency of the quasiparticle mass stay undamped until a time increasing with the length of the quenched interval, before eventually decaying as $t^{-1/2}$. The translation invariant case with no damping is recovered as the length of the interval goes to infinity.
\end{abstract}
\end{titlepage}

\newpage

\tableofcontents

\section{Introduction}
In a quantum quench an extended isolated system is in the ground state $|0\rangle$ of its Hamiltonian $H_0$ until the time $t=0$, when the sudden change of an interaction parameter leads to a new Hamiltonian $H$ that rules the unitary evolution for positive times. The quench provides the simplest way to produce nonequilibrium quantum dynamics, and is then the case in which the main theoretical questions can be addressed. The fact that the quantum state for $t>0$ is dynamically generated acting on a coupling eliminates ambiguities on the initialization of nonequilibrium evolution, but is the main source of theoretical difficulty. When the quasiparticle modes excited by the quench interact, which is almost always the case, the problem cannot be solved exactly \cite{quench} even in the one-dimensional case for which integrability yields so many exact results at equilibrium\footnote{The results of \cite{quench} on (un)solvability are obtained in the continuum and do not rule out some exception in lattice instances not admitting a continuum limit.}. However, a general study can be performed perturbatively in the quench parameter $\lambda$ ($\lambda=0$ amounts to no quench). For a translation invariant one-dimensional system with post-quench Hamiltonian $H=H_0+\lambda\int dx\,\Psi(x)$, the result for the one-point function of a local observable $\Phi(x)$ (e.g. the order parameter) at large times reads \cite{quench}
\EQ
\langle\Phi(x,t)\rangle=\langle\Phi\rangle_\lambda^\textrm{eq}+\lambda\left[\frac{2}{M^2}F_{1}^\Psi F_{1}^\Phi\,\cos Mt+O(t^{-3/2})\right]+O(\lambda^2)\,,
\label{hom1}
\EN
where $\langle\Phi\rangle_\lambda^\textrm{eq}$ is the equilibrium value in the theory with the Hamiltonian $H$ \cite{DV}, $M$ the quasiparticle mass, and $F_{1}^{\cal O}$ the matrix element of ${\cal O}$ between $|0\rangle$ and a  single-quasiparticle state; in case of several quasiparticle species, the term in the square bracket is summed over species. (\ref{hom1}) emerges as the most general analytic result available for long time evolution in nonequilibrium quantum dynamics. If we exclude that $|\langle\Phi(x,t)\rangle|$ indefinitely grows in time\footnote{If, in (\ref{hom1}), $g(t)$ is the sum of second and higher orders in $\lambda$, there are three possibilities for the large $t$ behavior: $g(t)$ goes to a constant; it oscillates; its absolute value indefinitely increases in time.}, (\ref{hom1}) shows the presence of undamped oscillations when the post-quench state includes a single-quasiparticle excitation mode ($F_{1}^\Psi\neq 0$) and the observable couples to this mode ($F_{1}^\Phi\neq 0$). The fact that the undamped oscillations require interacting quasiparticles (otherwise $\Psi$ creates only quasiparticle pairs and has $F_{1}^\Psi=0$) is one reason why they could be predicted only within the general theoretical formulation of \cite{quench}. 

In this paper we consider the case in which pre-quench translation invariance is broken by the quench. Although our results can be extended to more general patterns, we illustrate the formalism for quenches that change an interaction parameter in an interval $x\in[-b,b]$ of the system that extends along the whole $x$-axis. We then consider the post-quench Hamiltonian
\EQ
H=H_0+\lambda\int_{-b}^{b} dx\,\Psi(x)\,
\label{H}
\EN
and show that (\ref{hom1}) generalizes to
\EQ
\langle\Phi(x,t)\rangle=\langle\Phi(x)\rangle_\lambda^\textrm{eq}+\lambda\left[\frac{2}{\pi}F_{1}^\Psi F_{1}^\Phi\int dp\,\frac{\sin pb}{p}\,\frac{\cos(\sqrt{p^2+M^2}\,t+px)}{p^2+M^2}+O(t^{-2})\right]+O(\lambda^2)\,.
\label{inhom1}
\EN
The result implies, in particular, that for $F_{1}^\Psi F_{1}^\Phi\neq 0$ and $b\gg 1/M$ there are oscillations that at $x=0$ stay undamped up to a time increasing with $b$, before eventually decaying as $t^{-1/2}$. More generally, we show that the one-point function starts to appreciably differ from the pre-quench value only after a time $(|x|-b)/v_{\textrm{max}}$; $v_{\textrm{max}}$ is the maximal velocity of a quasiparticle and is $1$ in the units we adopt throughout the paper. For any finite $b$ the $t^{-1/2}$ damping occurs for any $x$ at sufficiently large times. 

In the next section we introduce the theoretical formalism, before analyzing in section~3 the time evolution of one-point functions and the oscillation pattern. The results are summarized and discussed in the last section.

\section{Quenches that break translation invariance}
\subsection{Post-quench state}
We consider an infinite one-dimensional system that before the quench is translation invariant and in the ground state $|0\rangle$ of the Hamiltonian $H_0$. We work in the basis of asymptotic quasiparticle states $|p_1,\ldots,p_n\rangle$ of the pre-quench theory, with $p_i$ denoting the momenta of the quasiparticles. For the sake of notational simplicity we refer to the case of a single quasiparticle species, generalizations being straightforward. The asymptotic states are eigenstates of $H_0$ with eigenvalues given by the sum of the quasiparticle energies $E_{i}=\sqrt{M^2+p_i^2}$. The quasiparticle mass $M$ measures the distance from a quantum critical point and is taken strictly positive. 

The quench at $t=0$ leads to the Hamiltonian (\ref{H}) and for $b$ finite breaks translation invariance. The quench excites quasiparticle modes and the pre-quench state $|0\rangle$ evolves into the state $|\psi_0\rangle=S_\lambda|0\rangle$, where 
\EQ
S_\lambda=T\,\exp\left(-i\lambda\int_0^\infty dt\int_{-b}^b dx\,\Psi(x,t)\right)
\label{Slambda}
\EN
($T$ denotes chronological ordering) is the operator whose matrix elements $\langle n|S_\lambda|0\rangle$ give the probability amplitude that the quench induces the transition $|0\rangle\to|n\rangle$. Here we are introducing the simplified notation $|n\rangle=|p_1,\ldots,p_n\rangle$. To first order in $\lambda$ we have
\EQ
|\psi_0\rangle\simeq |0\rangle+2\lambda\SumInt_{n,p_i}\,\frac{\sin Pb}{EP}\,[F_n^\Psi]^*\,|n\rangle\,,
\label{psi0}
\EN
where we defined $E=\sum_{i=1}^nE_i$, $P=\sum_{i=1}^nP_i$ and
\EQ
F_n^{\cal O}(p_1,\ldots,p_n)=\langle 0|{\cal O}(0,0)|p_1,\ldots,p_n\rangle\,,
\label{ff}
\EN
 introduced the notation
\EQ
\SumInt_{n,p_i}=\sum_{n=1}^\infty\frac{1}{n!}\int_{-\infty}^{\infty}\prod_{i=1}^n\frac{dp_i}{2\pi E_{i}}\,,
\label{sumint}
\EN
and used 
\EQ
\Psi(x,t)=e^{i{\cal P}x+iH_0t}\Psi(0,0)e^{-i{\cal P}x-iH_0t}\,,
\label{translation}
\EN
with ${\cal P}$ the momentum operator. As usual, an infinitesimal imaginary part given to the energy makes the time integral in (\ref{Slambda}) convergent\footnote{Notice that the sum in (\ref{sumint}) starts from $n=1$ rather than from $n=0$. The reason is that the $O(\lambda)$ contribution to (\ref{psi0}) with $n=0$ (which corresponds to $E=P=0$) diverges and must be subtracted. As recalled in \cite{quench} for the homogeneous case, such a term corresponds to vacuum energy renormalization and can be canceled introducing a counterterm in the Hamiltonian.}. It is worth emphasizing how (\ref{psi0}) provides the answer required within the field theoretical scattering framework, in which the quasiparticles cannot be followed in their time evolution and the post-quench state is expanded over the basis of asymptotic states, with coefficients determined by the integrated action (\ref{Slambda}) of the quench operator on the quenched interval for all positive times. The result shows that in the general case that we are considering the quench produces excitation modes with any number of quasiparticles and all possible momenta\footnote{A post-quench state with quasiparticles organized in pairs arises only when the quasiparticles do not interact before and after the quench, so that the Hamiltonians $H_0$ and $H$ are quadratic in the quasiparticle modes and yield $F_n^\Psi\propto\delta_{n,2}$. See \cite{CS} for an early study with noninteracting bosons and translation invariance broken before the quench.}.

\subsection{One-point functions}
A one-point function is determined by the expectation value of a local observable $\Phi$ on the post-quench state (5). As usual within the formalism of asymptotic states, the space-time dependence is carried by the operator and is extracted using (8) with $\Psi$ replaced by $\Phi$. The condition that the one-point function coincides at $t=0$ with its pre-quench value $\langle 0|\Phi(x,t)|0\rangle=\langle 0|\Phi(0,0)|0\rangle$ has to be separately required, there being no reason for it to be automatically fulfilled. As a consequence, the variation $\delta\langle\Phi(x,t)\rangle$ of the one-point function of a hermitian observable with respect to the pre-quench value is given, at first order in $\lambda$, by
\bea
\delta\langle\Phi(x,t)\rangle &\simeq &{\langle\psi_0|\Phi(x,t)|\psi_0\rangle-\langle 0|\Phi(0,0)|0\rangle}+C_\Phi(x)\nonumber\\
&\simeq & 4\lambda\,\SumInt_{n,p_j}\,\frac{\sin Pb}{EP}\,\mbox{Re}\{[F_n^\Psi]^*F_n^\Phi\,e^{-i(Et+Px)}\}+C_{\Phi}(x)\,,
\label{1point}
\eea
where we took into account that dividing by the normalization factor $\langle\psi_0|\psi_0\rangle=1+{O}(\lambda^2)$ is immaterial at first order, and added the term
\EQ
C_{\Phi}(x)\simeq -4\lambda\,\SumInt_{n,p_j}\,\frac{\sin Pb}{EP}\,\mbox{Re}\{[F_n^\Psi]^*F_n^\Phi\,e^{-iPx}\}
\label{C}
\EN
to impose the continuity at $t=0$, namely the condition $\delta\langle\Phi(x,0)\rangle=0$. 

In the next section we analyze the properties of the expression (\ref{1point}).

\section{Properties of the dynamics}
\subsection{Truncated light cone}
For $t$ large the exponential in (\ref{1point}) rapidly oscillates and suppresses the integrals over momenta unless the phase is stationary, i.e. unless 
\EQ
\partial_{p_j}[E_jt+p_j(x\pm b)]=v_jt+x\pm b=0\,,\hspace{1cm}j=1,2,\ldots,n\,,
\label{stationary}
\EN
where we introduced the velocities
\EQ
v_j=\frac{p_j}{\sqrt{M^2+p_j^2}}\,.
\EN
Since $v_j\in(-1,1)$, the stationarity condition (\ref{stationary}) is satisfied at large $t$ when 
\EQ
|x|<b+t\,.
\label{lightcone}
\EN
This result corresponds to the fact that the quench produces energy and momentum through the creation of the quasiparticles in the state (\ref{psi0}). These quasiparticles are produced at $t=0$ within the quenched interval $|x|<b$, but then propagate with maximal velocity $1$. As a consequence, the range of $|x|$ within which the one-point function appreciably differs from its pre-quench value is given by the truncated light cone (\ref{lightcone}). 

This derivation from first principles of the truncated light cone can be compared with that of the light cone associated to two-point functions in the translation invariant case \cite{lightcone}, in which the connectedness structure of matrix elements enters as an additional ingredient.

\subsection{Large time behavior}
\label{largetime}
For $x$ fixed and $t$ large enough, the stationarity condition $v_j=-(x\pm b)/t$ for the phase in (\ref{1point}) implies that the integrals receive a significant contribution only when all momenta $p_j$ are small, and then can be evaluated with $(\sin Pb)/P\to b$ and $E_j\to M+p_j^2/2M$. In addition, for particles with fermionic statistics, which is generic in interacting one-dimensional systems, the factor $[F_n^\Psi]^*F_n^\Phi$ in (\ref{1point}), evaluated for momenta all tending to zero, will be proportional to\footnote{The fermionic statistics requires that $F_n^{\cal O}$ vanishes when two momenta coincide. This leads, for each pair of momenta, to the presence of a factor $p_i-p_k$ that dominates the limit of small momenta. See e.g. \cite{review} for explicit illustrations of this behavior.} $\prod_{1\leq i<k \leq n}(p_i-p_k)^2$; this holds also for $n=1$, since for a scalar operator ${\cal O}$ the matrix element $F_1^{\cal O}$ is momentum-independent (it is also real in the cases of our interest). It is then straightforward to rescale the momenta and see that the $n$-quasiparticle contribution in (\ref{1point}) behaves at large times as $t^{-n^2/2}$. This yields the result in the square bracket of (\ref{inhom1}). As long as $F_1^\Psi F_1^\Phi\neq 0$, the leading contribution at large times is given by $n=1$ and is suppressed as $t^{-1/2}$; the first subleading contribution comes from $n=2$ and decays as $t^{-2}$. 

The result in the square bracket of (\ref{hom1}) for the translation invariant case is recovered from (\ref{1point}) using $\lim_{b\to\infty}(\sin Pb)/P=\pi\delta(P)$. When rescaling the momenta for large $t$, the factor $\delta(P)$ in the integrand yields an extra power $t^{1/2}$ with respect to the previous case, so that the $n$-quasiparticle contribution this time behaves as $t^{-(n^2-1)/2}$. Now the term $n=1$ is undamped\footnote{Notice that, in presence of several quasiparticle species $a=1,2,\ldots,k$, the term $n=1$ of the $O(\lambda)$ result for $|\psi_0\rangle$ gives to the one-point function an $O(\lambda^2)$ contribution $\sum_{a,b}\frac{\lambda^2F_{1,a}^\Psi F_{1,b}^\Psi}{(M_aM_b)^2}\langle p_a=0|\Phi(0)|p_b=0\rangle\,e^{i(M_a-M_b)t}$, which for $a\neq b$ corresponds to undamped oscillations with frequency $M_a-M_b$.}, while the term $n=2$ is suppressed as $t^{-3/2}$. 

It is worth observing that, while we study the general case of quasiparticles with a relativistic dispersion relation, the large time behavior is determined by the nonrelativistic limit of small momenta.

\subsection{Asymptotic offset}
\label{asympoffset}
In the {\it equilibrium} theory with Hamiltonian (\ref{H}) the first order contribution in $\lambda$ to a one-point function $\langle\Phi(x)\rangle_\lambda^{\textrm{eq}}$ is given by
\bea
\delta\langle\Phi(x)\rangle_\lambda^{\textrm{eq}} &\simeq & -i\lambda\int_{-\infty}^\infty dt\int_{-b}^b dy\,\langle 0|T\Psi(y,t)\Phi(x,0)|0\rangle_c
\label{deltaeq}\\
&=& -i\lambda\int_{-b}^b dy\left[\int_0^\infty dt\,\langle 0|\Psi(y,t)\Phi(x,0)|0\rangle_c+\int_{-\infty}^0 dt\,\langle 0|\Phi(x,0)\Psi(y,t)|0\rangle_c\right],
\nonumber
\eea
where the subscript $c$ indicates the connected part of the two-point function. Using (\ref{translation}) and expanding over asymptotic states we obtain
\bea
\delta\langle\Phi(x)\rangle_\lambda^{\textrm{eq}} &\simeq & -i\lambda\SumInt_{n,p_j}\int_{-b}^b dy\int_0^\infty dt\,e^{-iEt}\,2\textrm{Re}\{F_n^\Psi[F_n^\Phi]^*e^{iP(x-y)}\}\nonumber\\
&=& -4\lambda\SumInt_{n,p_j}\frac{\sin Pb}{EP}\,\textrm{Re}\{F_n^\Psi[F_n^\Phi]^*e^{iPx}\}\nonumber\\
&=& C_\Phi(x)\,,
\label{deltaeq2}
\eea
where we finally compared with (\ref{C}). Hence, recalling that the time-dependent part of (\ref{1point}) goes asymptotically to zero, we obtain
\EQ
\lim_{t\to\infty}\langle\Phi(x,t)\rangle=\langle\Phi(x)\rangle_\lambda^{\textrm{eq}}+O(\lambda^2)\,,
\label{offset}
\EN
as anticipated in (\ref{inhom1}). In the translation invariant case the r.h.s. of (\ref{offset}), which no longer depends on $x$, is the value around which the undamped oscillations occur (see (\ref{hom1})). 

\begin{figure}[t]
\begin{center}
\includegraphics[width=10cm]{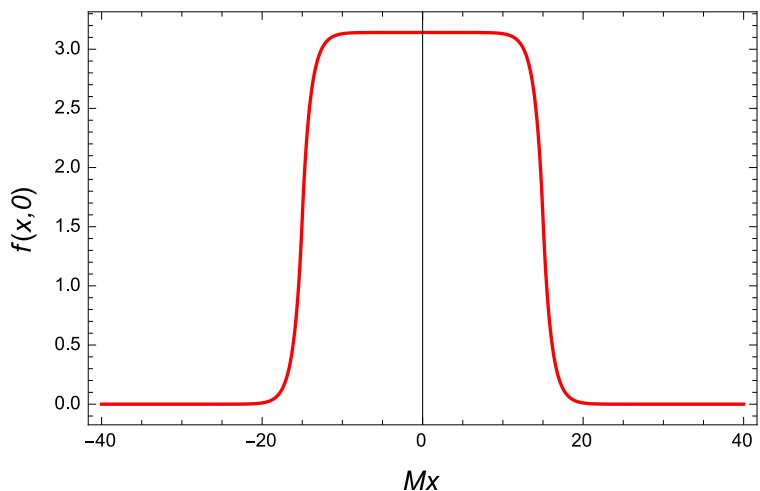}
\caption{The function (\ref{1part}) for $t=0$ and $Mb=15$. $f(x,0)$ approximates through (\ref{approx}) the equilibrium one-point function, and through (\ref{offset}) the large time limit of the nonequilibrium evolution.}
\label{oscill_offset}
\end{center}
\end{figure}

\subsection{Single-quasiparticle mode}
\label{single}
Let us rewrite (\ref{inhom1}) as 
\EQ
\langle\Phi(x,t)\rangle=\langle\Phi(x)\rangle_{\lambda}^\textrm{eq}+\lambda\left[\frac{2}{\pi M^2}F_{1}^\Psi F_{1}^\Phi\,f(x,t)+O(t^{-2})\right]+O(\lambda^2)\,,
\label{inhom2}
\EN
where
\bea
f(x,t) &=& M^2\int dp\,\frac{\sin pb}{p}\,\frac{\cos(\sqrt{p^2+M^2}\,t+px)}{p^2+M^2}\nonumber\\
&=& \int dq\,\frac{\sin qMb}{q}\,\frac{\cos(\sqrt{q^2+1}\,Mt+qMx)}{q^2+1}
\label{1part}
\eea
determines the contribution to time dependence coming from the single-quasiparticle mode. In the last expression we made explicit that $f(x,t)$ actually depends on the dimensionless quantities $Mx$, $Mt$ and $Mb$. For $F_{1}^\Psi F_{1}^\Phi\neq 0$ and for small quenches, i.e. for small $\lambda$, $f(x,t)$ determines the large time behavior of the one-point function until a time scale that goes to infinity as $\lambda$ is reduced. 

For short times, on the other hand, the analysis of section \ref{largetime} cannot be used to neglect the terms with $n>1$ in (\ref{1point}). It is known, however, that the contribution to the quasiparticle expansion is normally rapidly suppressed as $n$ increases (see \cite{DV} for the translation invariant case), so that $f(x,t)$ is expected to provide a good approximation also for short times. An interesting confirmation of this expectation comes from the fact that, together with (\ref{deltaeq2}) and (\ref{C}), it implies
\EQ
\delta\langle\Phi(x)\rangle_\lambda^{\textrm{eq}}\approx -\lambda\,\frac{2}{\pi M^2}\,F_{1}^\Psi F_{1}^\Phi\,f(x,0)
\label{approx}
\EN
for the first order contribution to the equilibrium expectation value. Figure \ref{oscill_offset} shows that the $n=1$ contribution is indeed sufficient to account for the expected result, namely that of a constant variation concentrated in the interval $|x|\lesssim b$. We recall that this result in turn determines through  (\ref{offset}) the large time limit of the nonequilibrium evolution. 

\begin{figure}[t]
\begin{center}
\includegraphics[width=9cm]{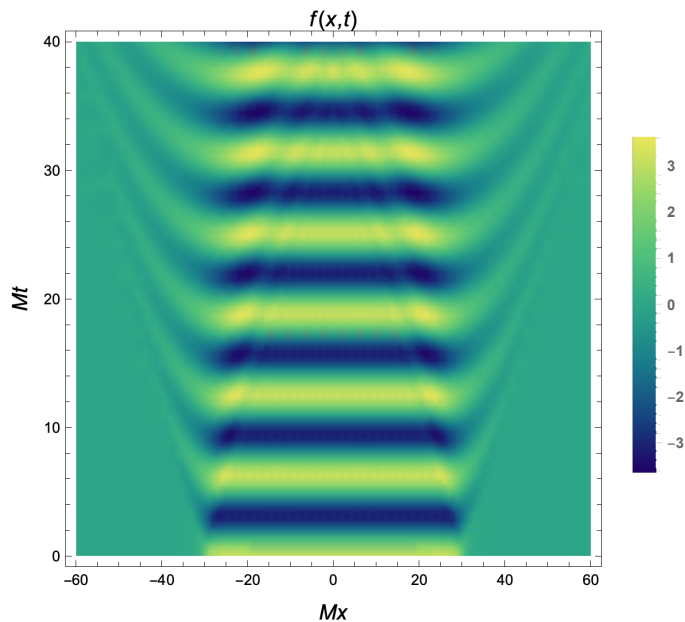}
\caption{The function $f(x,t)$ for $Mb=30$. When inserted in (\ref{inhom2}) the function determines the large time evolution of the one-point function, but provides a good approximation also for short times. The figure shows, in particular, that the time evolution takes place inside the truncated light cone $|x|<b+t$ (see also Figure~\ref{cone}).}
\label{dp}
\end{center}
\end{figure}

\begin{figure}
     \centering
     \begin{subfigure}[b]{7cm}
         \centering
         \includegraphics[width=\textwidth]{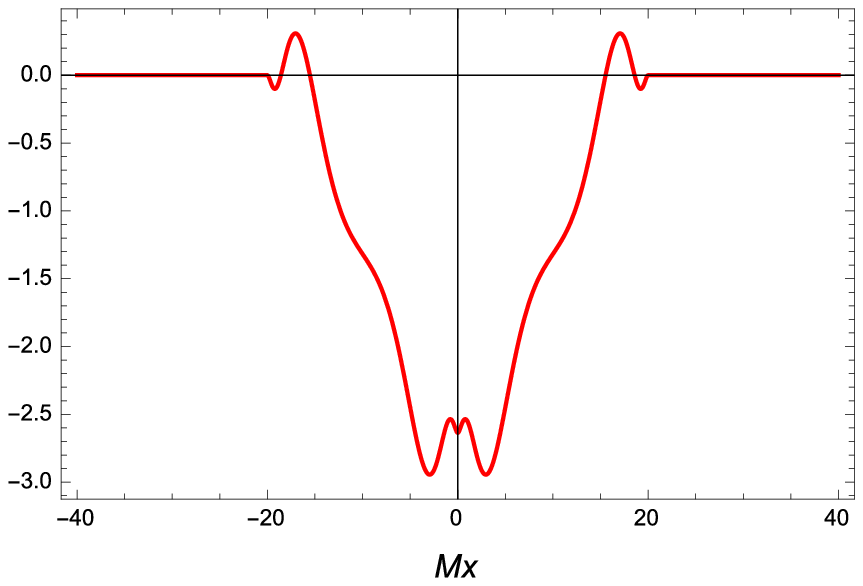}
        % \caption{$Mb=10$}
         %\label{fig:y equals x}
     \end{subfigure}
     \hfill
     \begin{subfigure}[b]{6.6cm}
         \centering
         \includegraphics[width=\textwidth]{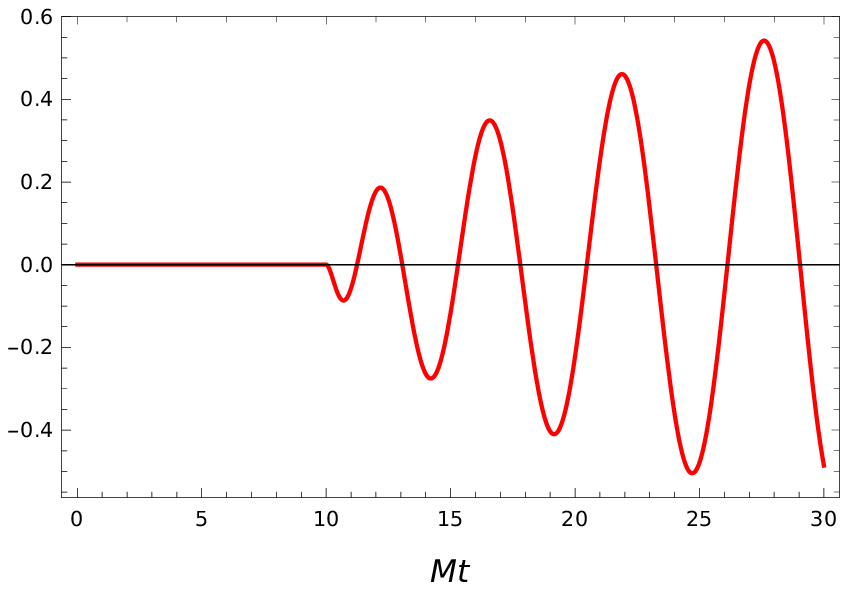}
        % \caption{$y=3sinx$}
         %\label{fig:three sin x}
     \end{subfigure}
\caption{The function $f(x,t)$ for $Mb=10$. {\it Left:} Mt=10. The function tends to zero outside the edges of the light cone located at $|x|=b+t$. {\it Right:} $Mx=20$. Time evolution becomes appreciable only after that the light cone is reached at time $t=|x|-b$.}
\label{cone}
\end{figure}

This discussion shows that the expression (\ref{inhom2}) (without the term $O(t^{-2})$) can be used not only for large times but, with good approximation, also for intermediate and short times. Hence, the function (\ref{1part}) allows a global view of the time evolution for small quenches. A plot of this function is shown in Figure~\ref{dp}. The truncated light cone is clearly visible and further illustrated in Figure~\ref{cone}. 

\begin{figure}[t]
\begin{center}
\includegraphics[width=10cm]{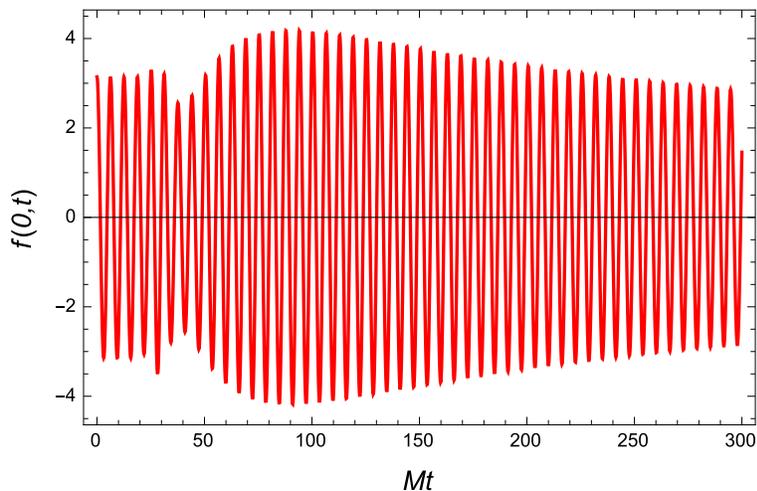}
\caption{The function $f(0,t)$ for $Mb=20$. There are undamped oscillations until $t\sim b$, while the $t^{-1/2}$ damping eventually sets in for larger times.}
\label{damping}
\end{center}
\end{figure}

Figure~\ref{dp} also exhibits the oscillations in time with frequency equal to the quasiparticle mass $M$. They are initially homogeneous in the interior of the quenched interval, but are progressively affected by the propagation inside the interval of the effects of translation invariance breaking at $|x|=b$. The evolution at $x=0$ is shown in Figure~\ref{damping}, where one sees that the oscillations stay undamped up to $t\sim b$, before eventually undergoing the $t^{-1/2}$ suppression derived in section~\ref{largetime}. Through this mechanism, absence of damping for all times in the translation invariant case is recovered as $b\to\infty$.

\section{Discussion}
We have extended the theory of quantum quenches introduced in \cite{quench} to the case of quenches that break the initial translation invariance of the system. We analyzed the case in which the one-dimensional homogeneous system is initially in the ground state and then undergoes an instantaneous quench (change of an interaction parameter) affecting only an interval of length $2b$. The theory determines the post-quench state and accounts, at the same time, for the variety of observable behaviors. Indeed, the excitation modes produced by the quench depend on the matrix elements $F^\Psi_n(p_1,\ldots,p_n)$ of the quench operator $\Psi$ on the quasiparticle states. In addition, the time evolution of a local observable $\Phi$ depends on its matrix elements $F^\Phi_n(p_1,\ldots,p_n)$. Remarkably, however, main properties of the dynamics can be determined in general. One of them is the truncated light cone that spreads away from the quenched interval as time increases. Another key feature is that for small quenches the long time behavior of one-point functions is determined by the lowest $n$, let us call it $n_0$, for which $F^\Psi_n F^\Phi_n\neq 0$. Since, due to the rapid convergence of the quasiparticle expansion, $n_0$ usually gives the main contribution also for intermediate and short times, a single term of the sum over $n$-quasiparticle excitations allows a global view of the time evolution. 

It is also remarkable that the theory unveils a {\it qualitative} difference between the case of noninteracting quasiparticles and the ordinary case in which the quasiparticles interact. Indeed, absence of interaction corresponds to a Hamiltonian quadratic in the quasiparticle modes, and then to $F_n^\Psi\propto\delta_{n,2}$. The interacting case, instead, gives $n_0=1$ unless $F_1^\Psi F_1^\Phi$ vanishes for symmetry reasons that are easy to check\footnote{See \cite{DV} for a series of examples in the translation invariant case. Considerations relying on internal symmetries hold unchanged also for the quenches that break translation invariance.}. It is the case $n_0=1$ that produces oscillations that stay undamped until a time increasing with $b$, and then for all times in the translation invariant case $b=\infty$. It was observed in \cite{quench} and illustrated in detail in \cite{DV} that this qualitative difference finds its simplest manifestation in the Ising spin chain, in which interaction among the quasiparticles is switched on by a longitudinal field. 

The general character of these features of one-dimensional nonequilibrium dynamics -- they are related to symmetry and presence of interaction -- explains that the theory reveals them already at first order in the expansion in the quench parameter $\lambda$. It is also clear that they do not depend on integrability\footnote{Integrability (of the pre-quench theory) becomes relevant if one wants exact expressions for the matrix elements $F_n^\Psi$ and $F_n^\Phi$ in (\ref{1point}), see \cite{DV,review}.}. The analytic results for $n_0=1$, which we illustrated in section~\ref{single}, are also largely universal for small quenches. In presence of several quasiparticle species there is a contribution from each species, and there are different oscillation frequencies for the different quasiparticle masses.  

The undamped oscillations of \cite{quench} for the translation invariant case have been observed, within the accessible time scales, in numerical and experimental studies (see \cite{BCH,RMCKT,KCTC,Lukin,Liu}). In \cite{BCH,Lukin,Liu} the oscillations were observed to arise only for special choices of the observable and of the initialization of the dynamics. The theory explains that this amounts to searching for the condition $F_1^\Psi F_1^\Phi\neq 0$. In \cite{RMCKT,KCTC} the actual quenches considered in the theory were implemented\footnote{Ref. \cite{KCTC} considers in particular the case in which a longitudinal field is switched on starting from the ferromagnetic phase of the Ising spin chain. In this case the single-quasiparticle modes responsible for undamped oscillations of the order parameter are generated nonperturbatively, through confinement (see \cite{review} for a review) of the topological excitations (kinks) of the ferromagnetic phase.}. The case of \cite{Liu} is particular for the fact that the oscillations, as well as results of \cite{lightcone} for the light cone spreading of correlations, were observed in a spin chain with long range interactions, thus indicating a remarkable robustness of the theory. 

The quenches considered in the present paper are also perfectly accessible to simulations and experiments and it would be interesting to check the predictions of the theory. The lack of translation invariance is a common feature with the experiment of \cite{KWW}, where integrability (of the translation invariant system at equilibrium) was suggested as a possible explanation for the appearance of persistent oscillations. As we already remarked, the theory does not assign to integrability a role in this matter. The simplest instance where to observe the pattern of Figure~\ref{dp} is that of the order parameter $\sigma$ in the Ising spin chain initially in the ground state of the paramagnetic phase, with the quench realized switching on a small longitudinal field. Our formulae then apply with $\Phi=\Psi=\sigma$ and $F_1^\sigma\neq 0$.


\begin{thebibliography}{99}
\bibitem{quench} G. Delfino, J. Phys. A 47 (2014) 402001.
\bibitem{DV} G. Delfino and J. Viti, J. Phys. A: Math. Theor. 50 (2017) 084004.
\bibitem{CS} S. Sotiriadis and J. Cardy,  J. Stat. Mech. (2008) P11003.
\bibitem{lightcone} G. Delfino, Phys. Rev. E 97, 062138 (2018).
\bibitem{review} G. Delfino, J. Phys. A 37 (2004) R45.
\bibitem{BCH} M.C. Banuls, J.I. Cirac and M.B. Hastings, Phys. Rev. Lett. 106 (2011) 050405.
\bibitem{RMCKT} T. Rakovszky, M. Mestyan, M. Collura, M. Kormos and G. Takacs,  Nucl. Phys. B 911, 805 (2016).
\bibitem{KCTC} M. Kormos, M. Collura, G. Takacs, and P. Calabrese, Nat. Phys. 13, 246 (2017).
\bibitem{Lukin} H. Bernien, S. Schwartz, A. Keesling, H. Levine, A. Omran, H. Pichler, S. Choi, A.S. Zibrov, M. Endres, M. Greiner, V. Vuletic and M.D. Lukin, Nature 551, 579 (2017).
\bibitem{Liu} F. Liu, R. Lundgren, P. Titum, G. Pagano, J. Zhang, C. Monroe and A.V. Gorshkov, Phys. Rev. Lett. 122, 150601 (2019).
\bibitem{KWW} T. Kinoshita, T. Wenger and D.S. Weiss, Nature 440, 900 (2006).
\end{thebibliography}
\end{document}